\documentclass[a4paper]{jpconf}
\usepackage{graphicx}
\usepackage{amsmath}
\expandafter\let\csname equation*\endcsname\relax

\expandafter\let\csname endequation*\endcsname\relax

\usepackage{amsmath}

\usepackage{graphicx}
\graphicspath{{Images/}}

\usepackage{tikz}
\usetikzlibrary{3d,calc}

\usetikzlibrary{datavisualization}
\usetikzlibrary{datavisualization.formats.functions}
\usetikzlibrary{snakes}

\usepackage{pgfplots}
\pgfplotsset{
   compat = 1.7 ,
   yticklabel style = {/pgf/number format/.cd,fixed,fixed zerofill,precision=4} ,
   samples = 100 ,
   scaled ticks = false 
}
\usepackage{pgfplots}
\usepackage{pgfplotstable}

\usepgfplotslibrary{fillbetween}

\usepgfplotslibrary{external} 
\begin{document}

\title[Krypton Bohr Hamiltonian Calculations]{Structure of Krypton Isotopes using  the Generalised Bohr Hamiltonian Method}

\author{David Muir${}^{1}$, Leszek Pr\'{o}chniak${}^{2}$, Alessandro Pastore${}^{1}$, Jacek Dobaczewski${}^{1,3,4}$}

\address{${}^{1}$Department of Physics, University of York, Heslington, York, Y010 5DD, United Kingdom \\
${}^{2}$Heavy Ion Laboratory, University of Warsaw, Warsaw, Poland\\
${}^{3}$Helsinki Institute of Physics, P.O. Box 64, 00014 University of Helsinki, Finland \\
${}^{4}$Institute of Theoretical Physics, Faculty of Physics, University of Warsaw, ul. Pasteura 5, PL-02093 Warsaw, Poland \\
}
\ead{davidmuir.physics@gmail.com}
\vspace{10pt}
\begin{indented}
\item[]\today
\end{indented}

\begin{abstract}
We investigate the properties of the excited spectra of the even-even isotopes of krypton using a  Generalised Bohr Hamiltonian  with three different Skyrme functionals.
In particular, we investigate the evolution of the low-lying $2^{+}_{1}$ and $4^{+}_{1}$ states and their associated electromagnetic transitions.
The model reproduces quite nicely the energy trends apart from ${}^{88}$Kr, where none of the interactions used here are able to grasp a sudden change in the energy spectrum.
Additionally, we explore the neutron deficient region ${}^{72-76}$Kr which is a proposed region for shape coexistence. We observe that the model can reproduce the structure of the experimental spectrum of ${}^{72}$Kr exceedingly well. 

\end{abstract}

\section{Introduction}

Nuclear Energy Density Functional (NEDF) is the tool of choice to describe properties of atomic nuclei from light to heavy and from proton to neutron drip-lines~\cite{ben03}. 
The current functionals have reached quite high accuracy in reproducing both ground state properties such as masses~\cite{gor09,gor09b}, as well as some features of the excited spectrum~\cite{ber07,sab07,sca13}.

The NEDF is typically formulated to describe nuclear properties in the intrinsic reference frame of the nucleus, thus allowing for the possibility of breaking symmetries like angular momentum or particle number~\cite{she19}. To compare the prediction of the model with the experimental measurement, one has to pass from the intrinsic to the laboratory frame where all these symmetries are restored.
A valid alternative to an explicit process of symmetry restoration~\cite{ben03}, is represented by the  Generalized Bohr Hamiltonian (GBH) ~\cite{boh52} method.
Since GBH is a scalar under rotations and, thus, its  eigenstates have good angular momentum. 

The GBH has been derived microscopically and linked to an underlying mean-field calculation. For more extensive reviews see for examples Refs~\cite{pro09,per14,mat16}.
Although the GBH takes into account only quadrupole degrees of freedom, it is well suited to describe the rotation-vibration coupling, which is important to describe the spectra of a large set of nuclei.
Compared to other more sophisticated methods such as the Generator Coordinate Method (GCM)~\cite{rin04}, the GBH has the enormous advantage of being able to use as input \emph{any} type of functional independently on the presence of density dependent terms or not~\cite{dug09} and it is thus free from issues related to self-interaction or poles.

 In this article, following the work done in Ref.~\cite{pro15}, we present the GBH method using, as a microscopic input, the NEDF calculations obtained using a family of Skyrme functionals~\cite{sky58}.  In particular, we examine the properties of the low-lying $2^{+}_{1}$ and $4^{+}_{1}$  states in krypton isotopes as well as their electromagnetic transitions.
These nuclei have received a lot of attention both theoretically and experimentally~\cite{cle07,gir09,rod14,fu13}, due to shape coexistance~\cite{hey11} of oblate and prolate deformations in their ground states, but also to a possible modification of the N=40 shell gaps around $^{76}$Kr.  
\section{Generalised Bohr Hamiltonian}
In order to describe  quadrupole collective excitations of nuclei, we use  the 5-dimensional Generalised Bohr Hamiltonian~\cite{row10}. The GBH is capable of describing mixtures of rotational and vibrational motions which arise in quadrupole deformed systems. It takes the form
\begin{align}\label{collective}
\hat{H}_{coll}&=\hat{T}_{vib}+\hat{T}_{rot}+\hat{V}\left(\beta,\gamma\right)\;,
\end{align}

\noindent where 
\begin{align}
\hat{T}_{\text{vib}}&=-\frac{1}{2\sqrt{wr}}\left\{\frac{1}{\beta^{4}}\left[\partial_{\beta}\left(\beta^{4}\sqrt{\frac{r}{w}}{B_{\gamma\gamma}}\right)\partial_{\beta}-\partial_{\beta}\left(\beta^{3}\sqrt{\frac{r}{w}}{B_{\beta\gamma}}\right)\partial_{\gamma}\right]\right. \nonumber\\
&\left. +\frac{1}{\beta\sin\left(3\gamma\right)}\left[-\partial_{\gamma}\left(\sqrt{\frac{r}{w}}\sin\left(3\gamma\right){B_{\beta\gamma}}\right)\partial_{\beta}+\frac{1}{\beta}\partial_{\gamma}\left(\sqrt{\frac{r}{w}}\sin\left(3\gamma\right){B_{\beta\beta}}\right)\partial_{\gamma}\right]\right\}\;, \\
\hat{T}_{\text{rot}}&=\frac{1}{2}\sum^{3}_{k=1}\frac{I^{2}_{k}\left(\Omega\right)}{4B_{k}\left(\beta,\gamma\right)\beta^{2}\sin^{2}\left(\gamma-2\pi k/3\right)}.
\end{align}
$\hat{V}\left(\beta,\gamma\right)$ is the potential energy of the nucleus and $I_k$  denotes the k-component of the angular momentum in the body-fixed frame of a
nucleus. All these quantities are parametrised in terms of the deformation parameters $\beta$ and $\gamma$ and of 6 mass parameters $B$.
In the previous expressions, we also used the shorthand notations $w=B_{\beta\beta}B_{\gamma\gamma}-B^{2}_{\beta\gamma}$ and $r=B_{x}B_{y}B_{z}.$

According to Ref.~\cite{pro15}, the potential energy surface $\hat{V}\left(\beta,\gamma\right)$ as well as the 6 mass parameters have been calculated solving the Hartree-Fock-Bogoliubov (HFB) equations for a set of parameters $\beta,\gamma$. In the present article, we use the HFODD solver~\cite{dob09} in Cartesian coordinates, to calculate the potential energy surface solving the constrained HFB equations at various $\beta,\gamma$ points.
The calculations are performed using a deformed harmonic oscillator basis with 16 major shells. We have checked that this value is rich enough to adequately converge both the energy and electromagnetic transitions of excited states. The mass parameters are calculated via the cranking approximation~\cite{gir79,bar11}.
To take into account the neglecting of time odd fields in such approximations, we uniformly rescale the 6 mass parameters by a factor of 1.3 in all following calculations. For more detailed discussion on the GBH we refer to Ref.~\cite{pro15}.

To perform the calculations we make use of 3 Skyrme functionals: UNEDF0~\cite{kor10}, UNEDF1~\cite{kor12} and UNEDF1$_{\text{SO}}$~\cite{shi14}. Since these functionals have been adjusted on open shell nuclei, the parameters of the pairing sector are fixed.
Although these functionals belong to the same family, they have been adjusted using slightly different fitting protocols and, as discussed below, they provide quite different deformation patterns in Kr isotopes.
\section{Results}
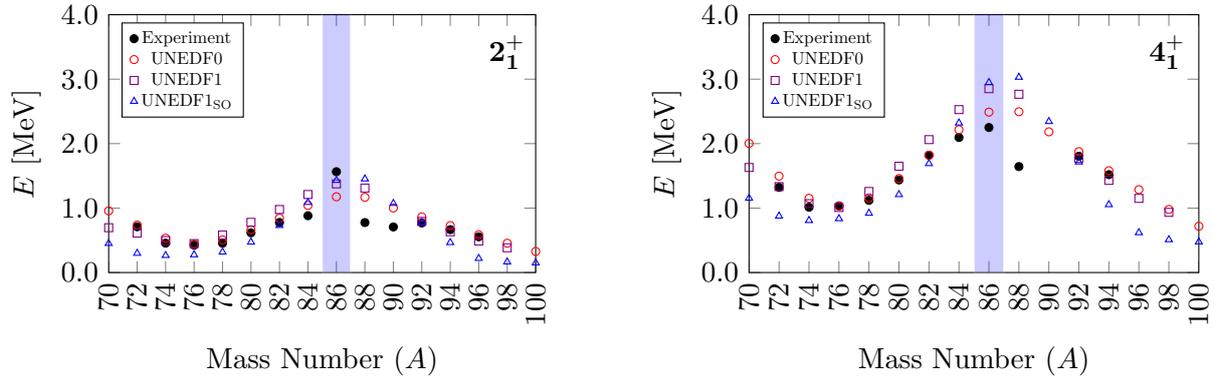
\begin{figure}[h!]
\centering
\begin{tikzpicture}
\def\PointSize{1.5pt}
\pgfplotsset{
   width = 7.2cm,
   height = 5cm, 
   compat = 1.9 ,
   xticklabel style = {rotate = 90},
   yticklabel style = {/pgf/number format/.cd,fixed,fixed zerofill,precision=1} ,
   samples=3300,
}
\begin{axis}
[
xtick distance = 2,
xmin = 70,
xmax = 100,
ymin = 0.0,
ymax = 4.0,
xlabel={Mass Number ($A$)},
ylabel={$E$ [MeV]},
legend pos=north west,
legend style = {nodes = {scale = 0.6, transform shape}}
]
\addplot[color = black, only marks,  mark size=\PointSize,mark=*,mark options={solid}] table{
72	0.70972
74	0.45561
76	0.42396
78	0.455033
80	0.6166
82	0.776526
84	0.881615
86	1.56461
88	0.77532
90	0.70713
92	0.7691
94	0.6655
96	0.5541
};
\addplot[color = red, only marks, mark = o, mark size=\PointSize] table{
70    0.9560752
72    0.7366201
74    0.5326215
76    0.4393176
78    0.5065395
80    0.6638374
82    0.8443295
84    1.0418014
86    1.1749851
88    1.1673227
90    1.0011564
92    0.8631399
94    0.7283160
96    0.5868754
98    0.4547765000
100   0.3273709000
};

\addplot[color = violet, only marks, mark = square, mark size=\PointSize] table{
  70  0.6954609000
  72  0.6115925000
  74  0.4953228000
  76  0.4487464000
  78  0.5830306000
  80  0.7784709000
  82  0.9773468000
  84  1.2104931000
  86  1.3753106000
  88  1.3106413000
  92  0.7935661000
  94  0.6323294000
  96  0.4891308000
  98  0.3840634000
};

\addplot[color = blue, only marks, mark = triangle, mark size=\PointSize] table{
  70  0.4513415000
  72  0.2978999000
  74  0.2647946000
  76  0.2753582000
  78  0.3181887000
  80  0.4717821000
  82  0.7302852000
  84  1.0877634000
  86  1.4296938000
  88  1.4524130000
  90  1.0725171000
  92  0.7783528000
  94  0.4650217000
  96  0.2190956000
  98  0.1639850000
 100  0.1506646000
};

\filldraw [draw = white, fill = blue, opacity = 0.2] (axis cs:85,0) rectangle (axis cs:87,5);

\node at (axis cs:96,3.0) [anchor=south west] {$\mathbf{2^{+}_{1}}$};
\legend{Experiment,UNEDF0, UNEDF1, UNEDF1$_{\text{SO}}$ }
\end{axis}
\end{tikzpicture}
\hfill
\begin{tikzpicture}
\def\PointSize{1.5pt}
\pgfplotsset{
   width = 7.5cm,
   height = 5cm, 
   compat = 1.9 ,
   xticklabel style = {rotate = 90},
   yticklabel style = {/pgf/number format/.cd,fixed,fixed zerofill,precision=1} ,
   samples=3300,
}
\begin{axis}
[
xtick distance = 2,
xmin = 70,
xmax = 100,
ymin = 0.0,
ymax = 4.0,
xlabel={Mass Number ($A$)},
ylabel={$E$ [MeV]},
legend pos=north west,
legend style = {nodes = {scale = 0.6, transform shape}}
]
\addplot[color = black, only marks, mark size=\PointSize] table{
72	1.3214
74	1.01332
76	1.03462
78	1.11948
80	1.43609
82	1.820536
84	2.095
86	2.25001
88	1.64378
92	1.804
94	1.5187

};
\addplot[color = red, only marks, mark = o, mark size=\PointSize] table{
70    2.0034901
72    1.4957642
74    1.1511392
76    1.0301761
78    1.1585619
80    1.4587794
82    1.8210786
84    2.2161085
86    2.4884507
88    2.4949559
90    2.1832906
92    1.8764014
94    1.5800383
96    1.2862074
98    0.9790003
100   0.7184858
};

\addplot[color = violet, only marks, mark = square, mark size=\PointSize] table{
  70  1.6299892000
  72  1.3332155000
  74  1.0689834000
  76  1.0088696000
  78  1.2572871000
  80  1.6497157000
  82  2.0623145000
  84  2.5280070000
  86  2.8526147000
  88  2.7659438000
  92  1.7715227000
  94  1.4301765000
  96  1.1506678000
  98  0.9336680000
};

\addplot[color = blue, only marks, mark = triangle, mark size=\PointSize] table{
  70  1.1537256000
  72  0.8761063000
  74  0.8071674000
  76  0.8355899000
  78  0.9199542000
  80  1.2092103000
  82  1.6898507000
  84  2.3203595000
  86  2.9486410000
  88  3.0296513000
  90  2.3427413000
  92  1.7151443000
  94  1.0531508000
  96  0.6183401000
  98  0.5092291000
 100  0.4754178000
};
\filldraw [draw = white, fill = blue, opacity = 0.2] (axis cs:85,0) rectangle (axis cs:87,5);

\node at (axis cs:96,3.0) [anchor=south west] {$\mathbf{4^{+}_{1}}$};
\legend{Experiment, UNEDF0 , UNEDF1, UNEDF1$_{\text{SO}}$}
\end{axis}
\end{tikzpicture}
\caption{(Colour Online) We represent the evolution of the energies expressed in MeV of the first excited $2^{+}_{1}$  (left panel) and $4^{+}_{1}$ (right panel)  states across the krypton isotopic chain for the 3 functionals: UNEDF0 represented by hollow circles; UNEDF1 represented by hollow squares and UNEDF1$_{\text{SO}}$ represented by hollow triangles and compare them to the experimental results~\cite{DATA} represented by the solid dots. The light blue bands highlight the semi-magic ${}^{86}$Kr nucleus.
}
\label{2+ and 4+ States of the Krypton Isotopic Chain}
\end{figure}
\begin{figure}[h]
\centering
\begin{tikzpicture}
\def\ErrorBandColour{orange}
\def\PointSize{  1.5000000000pt}
\pgfplotsset{
width =   6.5000000000cm,
height =   6.5000000000cm,
compat = 1.9 ,
xticklabel style = {/pgf/number format/.cd,fixed,fixed zerofill,precision=1} ,
yticklabel style = {/pgf/number format/.cd,fixed,fixed zerofill,precision=1} ,
samples=3300,
}
\begin{axis}
[
xtick distance = 0.5,
ytick distance = 0.5,
xmin =   0.0000000000,
xmax =   3.5000000000,
ymin =   0.0000000000,
ymax =   3.5000000000,
xlabel={Experimental Energy $E$ [MeV]},
ylabel={Theory Energy $E$ [MeV]},
legend pos=south east
]
\addplot[color = \ErrorBandColour, thick] {x};
\addplot[name path = ERRORA, color = \ErrorBandColour!70] {x+0.1};
\addplot[name path = ERRORB, color = \ErrorBandColour!70] {x-0.1};
\addplot[name path = ERRORC, color = \ErrorBandColour!40] {x+0.2};
\addplot[name path = ERRORD, color = \ErrorBandColour!40] {x-0.2};
\addplot [\ErrorBandColour!80, opacity = 0.6] fill between [of = ERRORA and ERRORB];
\addplot [\ErrorBandColour!40, opacity = 0.6] fill between [of = ERRORC and ERRORD];
\addplot[color = blue, opacity = 0.3, only marks, mark size=3.0*\PointSize] table{
   1.5646100000000001        1.1749851000000000 
   1.5646100000000001        1.3753105999999999   
   1.5646100000000001        1.4296937999999999     
};
\addplot[color = red, only marks, mark = o, mark size=\PointSize] table{
  0.70972000000000002       0.73662010000000000     
  0.45561000000000001       0.53262149999999997     
  0.42396000000000000       0.43931759999999997     
  0.45503300000000002       0.50653950000000003     
  0.61660000000000004       0.66383740000000002     
  0.77652600000000005       0.84432949999999996     
  0.88161500000000004        1.0418014000000000     
   1.5646100000000001        1.1749851000000000     
  0.77532000000000001        1.1673226999999999     
  0.70713000000000004        1.0011563999999999     
  0.76910000000000001       0.86313989999999996     
  0.66549999999999998       0.72831599999999996     
  0.55410000000000004       0.58687540000000005     
};
\addplot[color = violet, only marks, mark = square, mark size=\PointSize] table{
  0.70972000000000002       0.61159249999999998     
  0.45561000000000001       0.49532280000000001     
  0.42396000000000000       0.44874639999999999     
  0.45503300000000002       0.58303059999999995     
  0.61660000000000004       0.77847089999999997     
  0.77652600000000005       0.97734679999999996     
  0.88161500000000004        1.2104931000000001     
   1.5646100000000001        1.3753105999999999     
  0.77532000000000001        1.3106412999999999     
  0.76910000000000001       0.79356610000000005     
  0.66549999999999998       0.63232940000000004     
  0.55410000000000004       0.48913079999999998     
};
\addplot[color = blue, only marks, mark = triangle, mark size=\PointSize] table{
  0.70972000000000002       0.29789990000000000     
  0.45561000000000001       0.26479459999999999     
  0.42396000000000000       0.27535820000000000     
  0.45503300000000002       0.31818869999999999     
  0.61660000000000004       0.47178209999999998     
  0.77652600000000005       0.73028519999999997     
  0.88161500000000004        1.0877634000000000     
   1.5646100000000001        1.4296937999999999     
  0.77532000000000001        1.4524130000000000     
  0.70713000000000004        1.0725171000000000     
  0.76910000000000001       0.77835279999999996     
  0.66549999999999998       0.46502169999999998     
  0.55410000000000004       0.21909560000000000     
};
\node [color = black] at (axis cs:0.3,3.2) {$\mathbf{2^{+}_{1}}$};
\end{axis}
\end{tikzpicture}
\hfill
\begin{tikzpicture}
\def\ErrorBandColour{orange}
\def\PointSize{  1.5000000000pt}
\pgfplotsset{
width =   6.5000000000cm,
height =   6.5000000000cm,
compat = 1.9 ,
xticklabel style = {/pgf/number format/.cd,fixed,fixed zerofill,precision=1} ,
yticklabel style = {/pgf/number format/.cd,fixed,fixed zerofill,precision=1} ,
samples=3300,
}
\begin{axis}
[
xtick distance = 0.5,
ytick distance = 0.5,
xmin =   0.0000000000,
xmax =   3.5000000000,
ymin =   0.0000000000,
ymax =   3.5000000000,
xlabel={Experimental Energy $E$ [MeV]},
ylabel={Theory Energy $E$ [MeV]},
legend pos=north west
]
\addplot[color = \ErrorBandColour, thick] {x};
\addplot[name path = ERRORA, color = \ErrorBandColour!70] {x+0.1};
\addplot[name path = ERRORB, color = \ErrorBandColour!70] {x-0.1};
\addplot[name path = ERRORC, color = \ErrorBandColour!40] {x+0.2};
\addplot[name path = ERRORD, color = \ErrorBandColour!40] {x-0.2};
\addplot [\ErrorBandColour!80, opacity = 0.6] fill between [of = ERRORA and ERRORB];
\addplot [\ErrorBandColour!40, opacity = 0.6] fill between [of = ERRORC and ERRORD];
\addplot[color = blue, opacity = 0.3, only marks, mark size=3.0*\PointSize] table{
   2.2500100000000001        2.4884507000000000 
   2.2500100000000001        2.8526147000000002 
   2.2500100000000001        2.9486409999999998     
};
\addplot[color = red, only marks, mark = o, mark size=\PointSize] table{
   1.3213999999999999        1.4957642000000000     
   1.0133200000000000        1.1511392000000000     
   1.0346200000000001        1.0301761000000000     
   1.1194800000000000        1.1585619000000000     
   1.4360900000000001        1.4587794000000001     
   1.8205359999999999        1.8210786000000001     
   2.0950000000000002        2.2161084999999998     
   2.2500100000000001        2.4884507000000000     
   1.6437800000000000        2.4949558999999999     
   1.8040000000000000        1.8764014000000000     
   1.5186999999999999        1.5800383000000000     
};
\addplot[color = violet, only marks, mark = square, mark size=\PointSize] table{
   1.3213999999999999        1.3332155000000001     
   1.0133200000000000        1.0689834000000000     
   1.0346200000000001        1.0088695999999999     
   1.1194800000000000        1.2572871000000001     
   1.4360900000000001        1.6497157000000000     
   1.8205359999999999        2.0623144999999998     
   2.0950000000000002        2.5280070000000001     
   2.2500100000000001        2.8526147000000002     
   1.6437800000000000        2.7659438000000001     
   1.8040000000000000        1.7715227000000000     
   1.5186999999999999        1.4301765000000000     
};
\addplot[color = blue, only marks, mark = triangle, mark size=\PointSize] table{
   1.3213999999999999       0.87610630000000000     
   1.0133200000000000       0.80716739999999998     
   1.0346200000000001       0.83558990000000000     
   1.1194800000000000       0.91995420000000006     
   1.4360900000000001        1.2092103000000001     
   1.8205359999999999        1.6898507000000000     
   2.0950000000000002        2.3203594999999999     
   2.2500100000000001        2.9486409999999998     
   1.6437800000000000        3.0296512999999998     
   1.8040000000000000        1.7151443000000000     
   1.5186999999999999        1.0531508000000001     
};
\node [color = black] at (axis cs:0.3,3.2) {$\mathbf{4^{+}_{1}}$};
\end{axis}
\end{tikzpicture}
\caption{(Colour Online) Shows the first $2^{+}_{1}$ (left panel) and $4^{+}_{1}$ (right panel) theoretical energy states against the corresponding experimentally measured states~\cite{DATA} for all 3 UNEDF functionals represented by hollow: circles for UNEDF0; squares for UNEDF1 and triangles for UNEDF1$_{\text{SO}}$. To guide the eye we add errorbars of $\pm 100$ keV (inner bound) and $\pm 200$ keV (outer bound). Highlighted in large circles are the semi-magic ${}^{86}$Kr nuclei.}
\label{Kr Theory vs Experiment}
\end{figure}
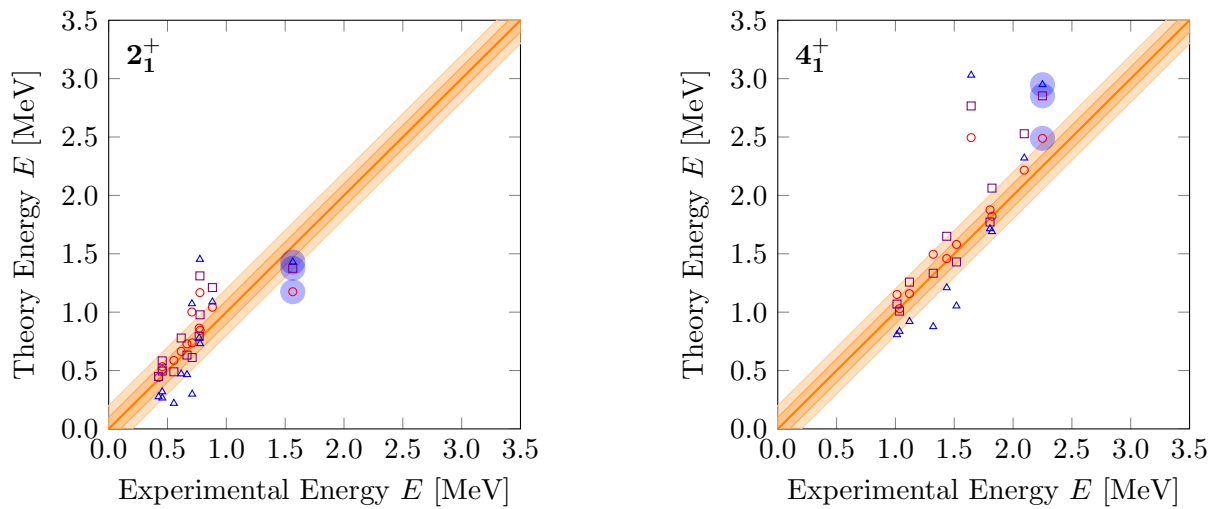
In Fig. \ref{2+ and 4+ States of the Krypton Isotopic Chain}, we show the evolution of energies of the  $2_1^+$ and $4_1^+$ states along the Kr isotopic chain obtained for the different functionals.
 We observe that the trend of the  $2_1^+$ is fairly well reproduced by all functionals, with an energy increase around the shell closure at $N=50$. A similar good agreement was also shown in Refs~\cite{gir09,fu13} also based on similar GBH methods but using different functionals.
The  evolution of $4_1^+$ states is also fairly well reproduced in the region $^{74-84}$Kr, but beyond the shell closure we observe the presence of a possible staggering in $^{88}$Kr and $^{92}$Kr which deviates quite remarkably from the smooth trend predicted by the GBH model.
 
 To better quantify the agreement between theory and experiment, we have plotted in Fig.\ref{Kr Theory vs Experiment} the theoretical energies of the  $2_1^+$  (left panel) and $4_1^+$  (right panel) as a function of the experimental ones.
 In the case of perfect match theory/experiment the points should lie on the diagonal line.
As already seen in Fig. \ref{2+ and 4+ States of the Krypton Isotopic Chain}, the energies of the $2^+_1$ are fairly well reproduced by UNEDF0 and UNEDF1, with most of the points falling within $\pm$200 keV from the experimental value. This value has not been obtained by any rigorous error analysis~\cite{dob14} and it should be considered only as a visual help for the reader to estimate the scattering of the data.
The results obtained with UNEDF1$_{\text{SO}}$ manifest somehow larger discrepancies, in particular it has the tendency to systematically underestimate the energy of the $2^+_1$ state.
 Similar conclusion holds for the $4^+_1$, we observe that the UNEDF0 data are most of the time quite close to the experimental data.
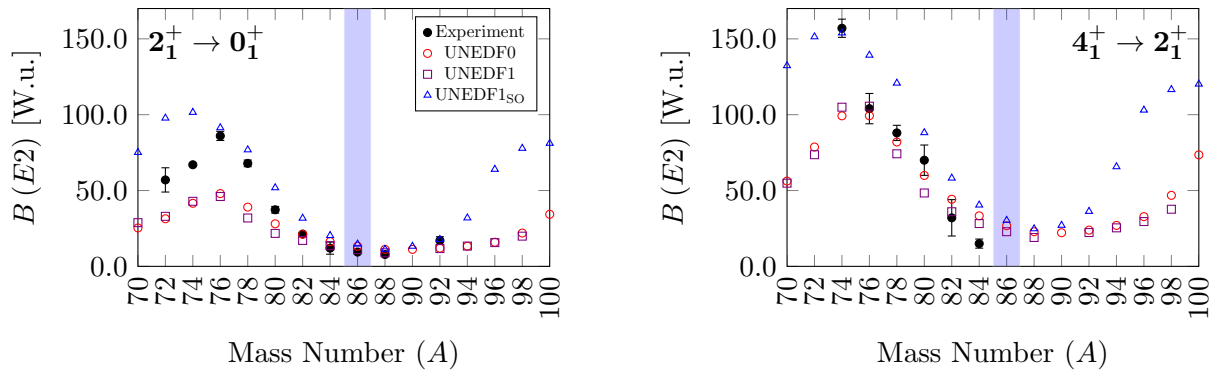
\begin{figure}[h!]
\centering
\begin{tikzpicture}
\def\PointSize{1.5pt}
\pgfplotsset{
   width = 7.0cm,
   height = 5cm, 
   compat = 1.9 ,
   xticklabel style = {rotate = 90},
   yticklabel style = {/pgf/number format/.cd,fixed,fixed zerofill,precision=1} ,
   samples=3300,
}
\begin{axis}
[
xtick distance = 2,
xmin = 70,
xmax = 100,
ymin = 0.0,
ymax = 170.0,
xlabel={Mass Number ($A$)},
ylabel={$B\left(E2\right)$ [W.u.]},
legend pos=north east,
legend style = {nodes = {scale = 0.6, transform shape}}
]

\addplot [color=black, only marks, mark size=\PointSize]
plot [error bars/.cd, y dir = both, y explicit] 
table [x=A, y=BE2W, y error = Error] {UNEDF0_BE2W_2_0.txt};

\addplot[color = red, only marks, mark = o, mark size=\PointSize] table{
70	25.311
72	31.385
74	41.583
76	47.98
78	39.048
80	28.061
82	21.165
84	16.328
86	13.258
88  11.180
90	11.198
92	11.988
94  13.313
96	15.904
98  21.960
100 34.304
};

\addplot[color = violet, only marks, mark = square, mark size=\PointSize] table{
70	28.827
72	32.976
74	42.895
76	46.066
78	31.883
80	21.73
82	17.092
84	13.657
86	11.07
88	8.96
92	11.731
94	13.375
96	15.655
98	19.917
};
\addplot[color = blue, only marks, mark = triangle, mark size=\PointSize] table{
70	75.122
72	97.764
74	101.524
76	91.4
78	76.828
80	51.825
82	31.756
84	20.341
86	14.704
88	11.223
90  13.256
92	17.722
94	31.934
96	63.981
98	77.826
100	81.142
};

\filldraw [draw = white, fill = blue, opacity = 0.2] (axis cs:85,0) rectangle (axis cs:87,170);

\node at (axis cs:70,130.0) [anchor=south west] {$\mathbf{2^{+}_{1}}\rightarrow \mathbf{0^{+}_{1}}$};
\legend{Experiment,UNEDF0, UNEDF1, UNEDF1$_{\text{SO}}$}
\end{axis}
\end{tikzpicture}
\hfill
\begin{tikzpicture}
\def\PointSize{1.5pt}
\pgfplotsset{
   width = 7.0cm,
   height = 5cm, 
   compat = 1.9 ,
   xticklabel style = {rotate = 90},
   yticklabel style = {/pgf/number format/.cd,fixed,fixed zerofill,precision=1} ,
   samples=3300,
}
\begin{axis}
[
xtick distance = 2,
xmin = 70,
xmax = 100,
ymin = 0.0,
ymax = 170.0,
xlabel={Mass Number ($A$)},
ylabel={$B\left(E2\right)$ [W.u.]},
legend pos=north west,
legend style = {nodes = {scale = 0.6, transform shape}}
]
\addplot [color=black, only marks, mark size=\PointSize]
plot [error bars/.cd, y dir = both, y explicit] 
table [x=A, y=BE2W, y error = Error] {UNEDF0_BE2W_4_2.txt};

\addplot[color = red, only marks, mark = o, mark size=\PointSize] table{
70	56.209
72	78.68
74	99.241
76	99.317
78	81.997
80	59.918
82	44.25
84	33.306
86	26.784
88  22.828
90	22.181
92	23.878
94  27.003
96	32.78
98  46.822
100 73.512
};

\addplot[color = violet, only marks, mark = square, mark size=\PointSize] table{
70	54.833
72	73.619
74	104.938
76	105.62
78	74.234
80	48.382
82	36.013
84	28.217
86	22.854
88	19.039
92	22.301
94	25.398
96	29.552
98	37.663
};
\addplot[color = blue, only marks, mark = triangle, mark size=\PointSize] table{
70	132.313
72	151.363
74	153.894
76	139.241
78	120.809
80	88.141
82	58.197
84	40.429
86	30.332
88	24.774
90  27.078
92	36.256
94	65.695
96	103.027
98	116.556
100	120.185
};
\filldraw [draw = white, fill = blue, opacity = 0.2] (axis cs:85,0) rectangle (axis cs:87,170);

\node at (axis cs:90,130.0) [anchor=south west] {$\mathbf{4^{+}_{1}}\rightarrow \mathbf{2^{+}_{1}}$};
\end{axis}
\end{tikzpicture}
\caption{(Colour Online) Shows the $B\left(E2:2^{+}_{1}\rightarrow 0^{+}_{1}\right)$ (left panel) and $B\left(E2:4^{+}_{1}\rightarrow 2^{+}_{1}\right)$ (right panel) transition probabilities~\cite{DATA} across the Krypton isotopic chain for the 3 UNEDF functionals represented by hollow: circles for UNEDF0; squares for UNEDF1 and triangles for UNEDF1$_{\text{SO}}$.
}
\label{2+ and 4+ BE2W of the Krypton Isotopic Chain}
\end{figure}

\begin{figure}[h]
\centering
%
%
\includegraphics[width=0.29\textwidth]{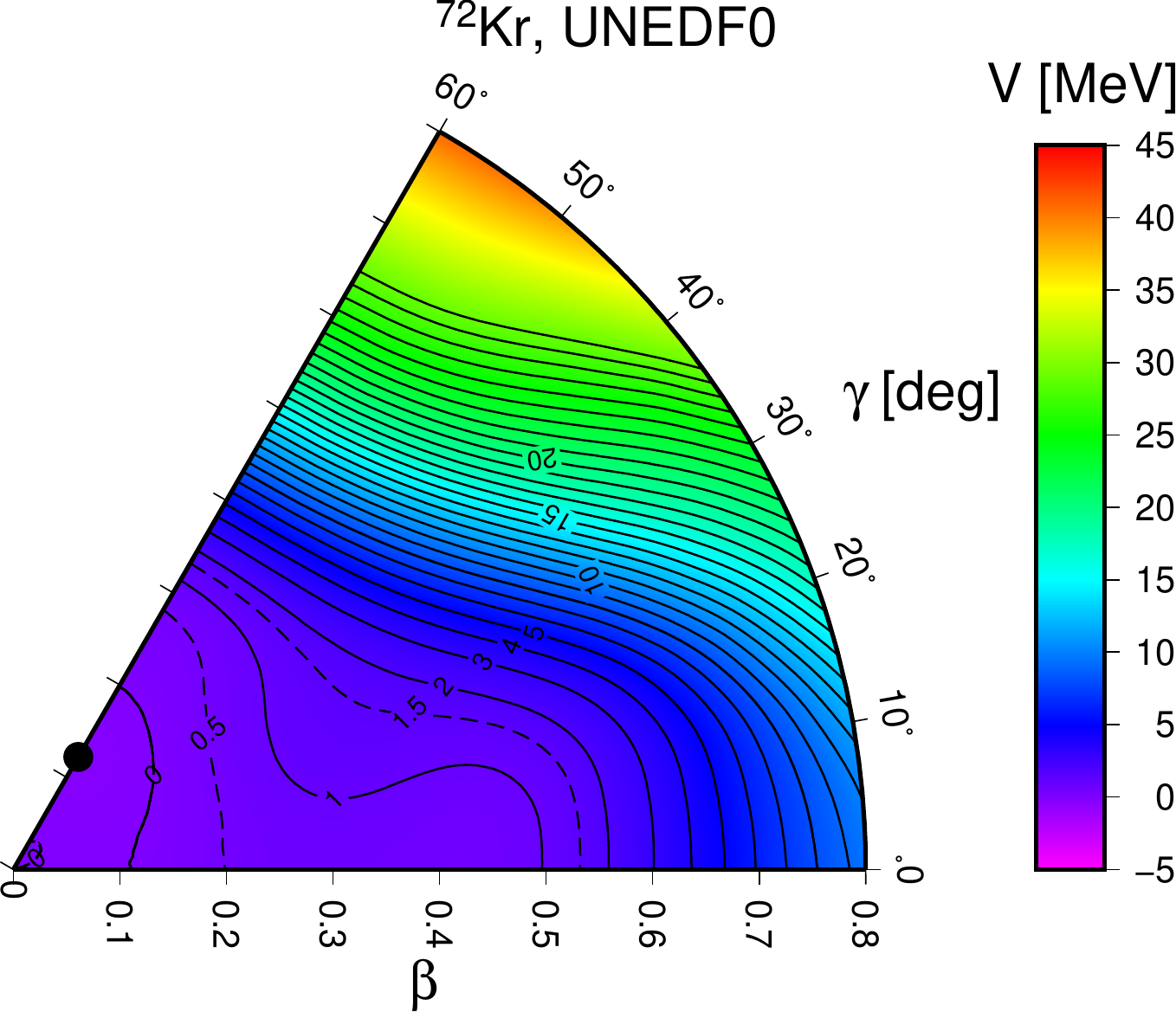}
\hfill
\includegraphics[width=0.29\textwidth]{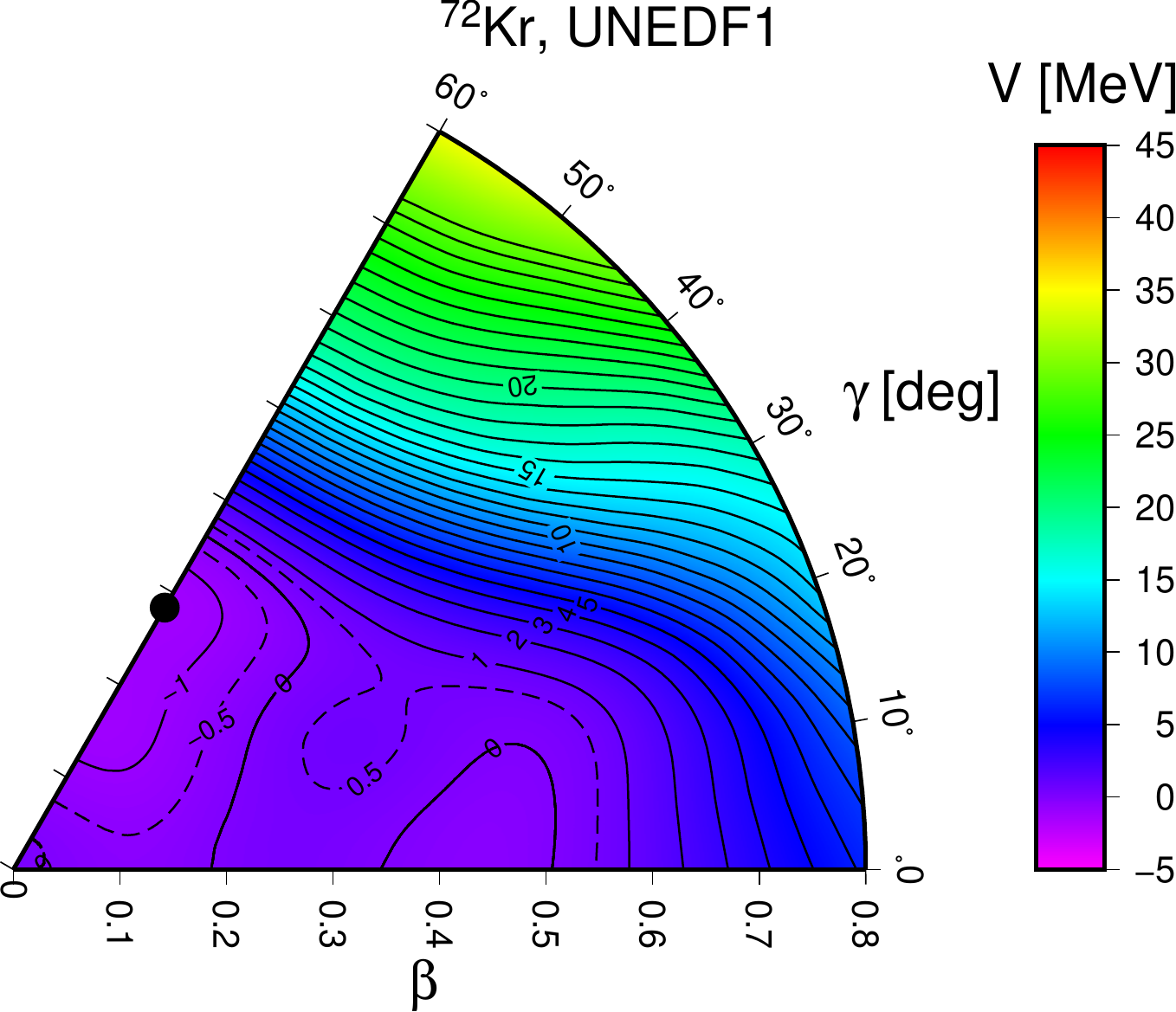}
\hfill
\includegraphics[width=0.29\textwidth]{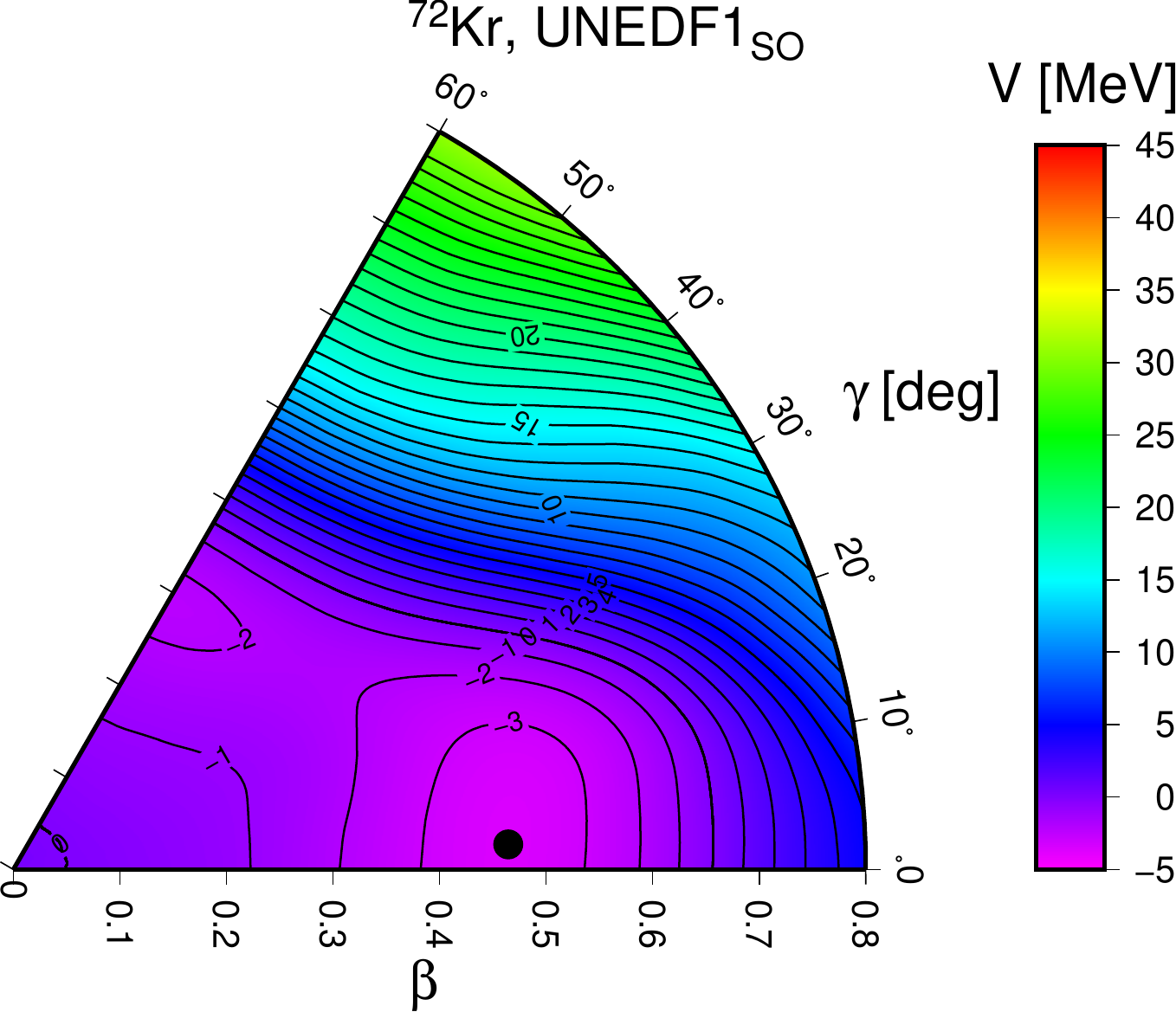}

\includegraphics[width=0.29\textwidth]{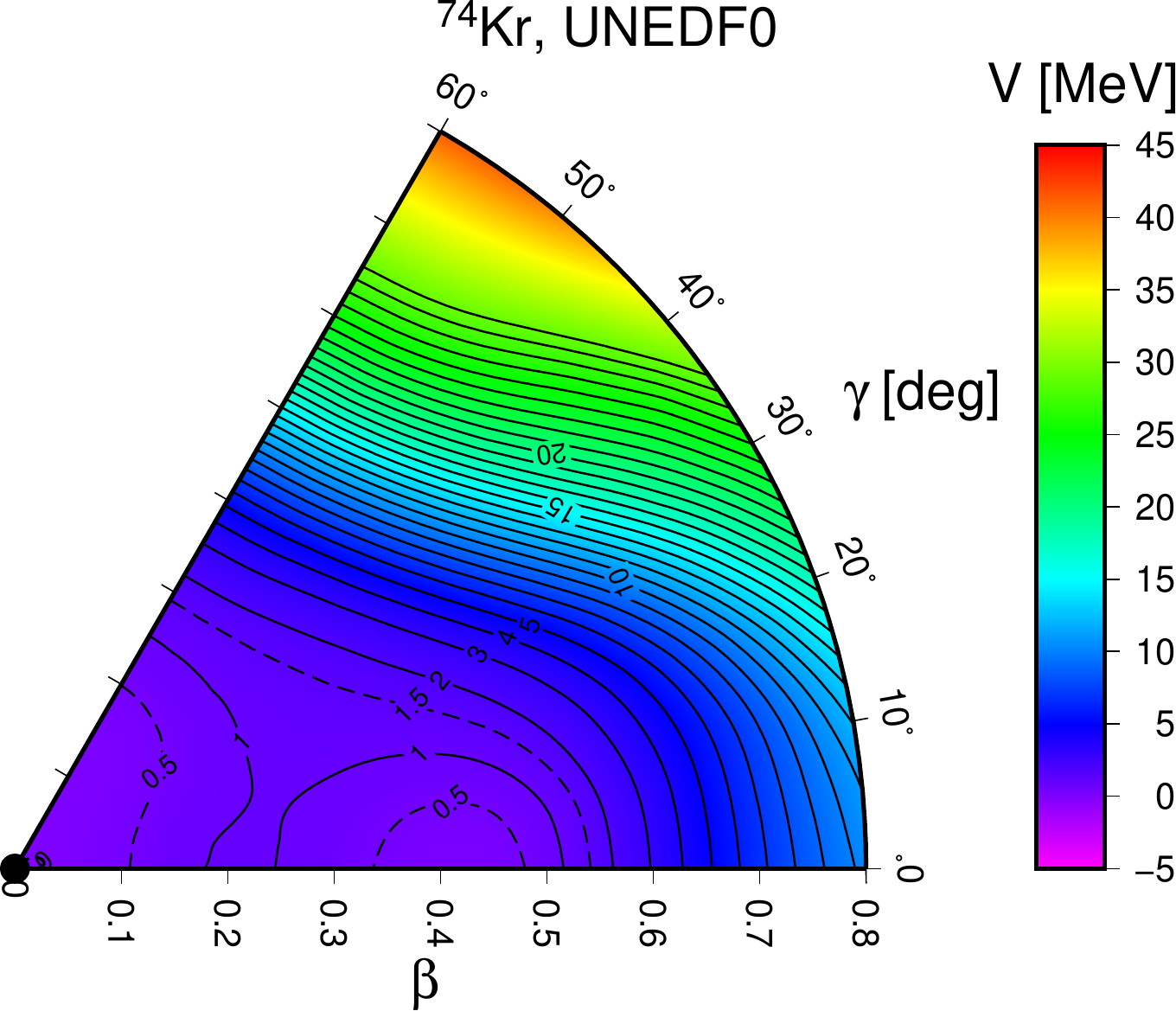}
\hfill
\includegraphics[width=0.29\textwidth]{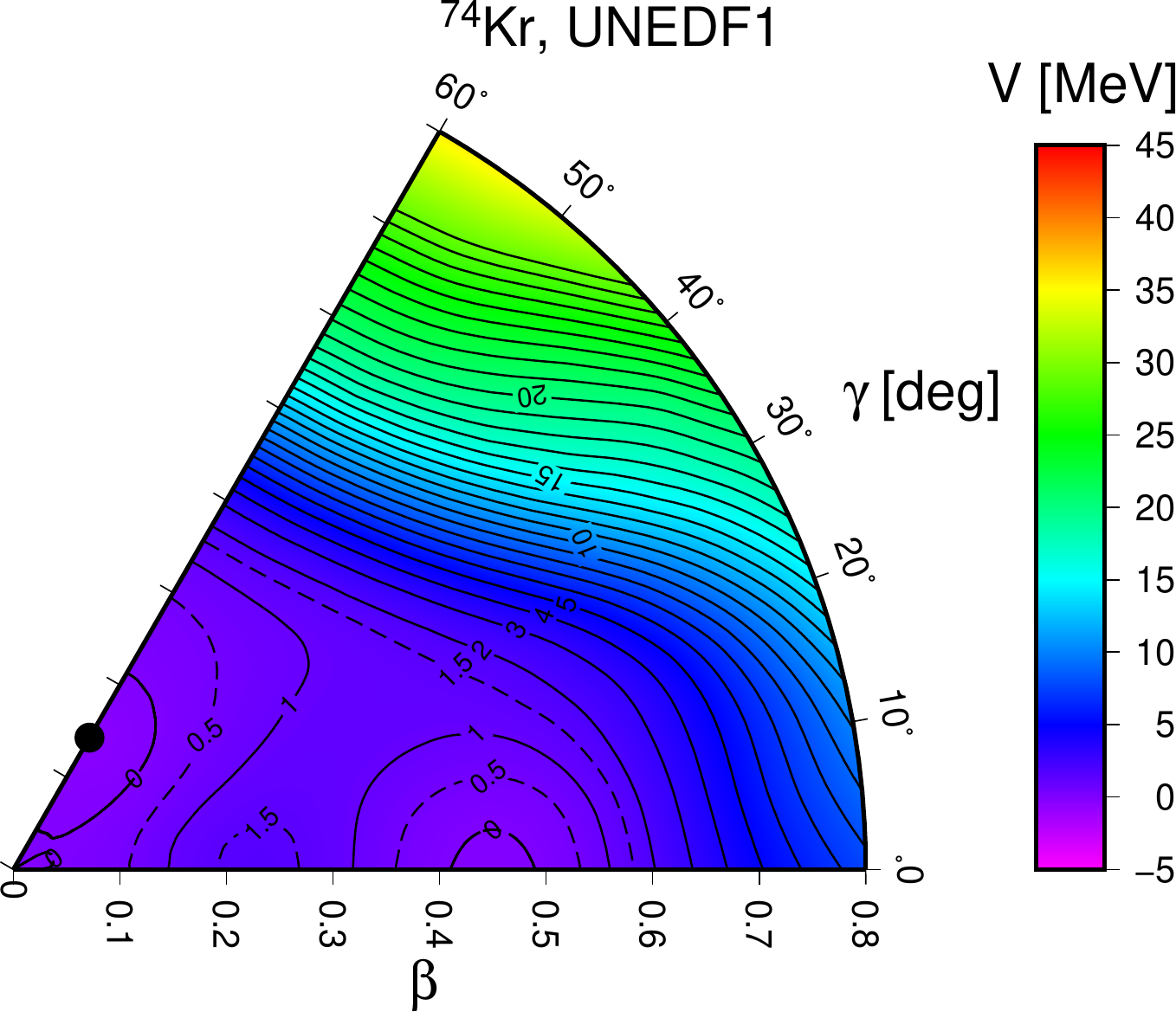}
\hfill
\includegraphics[width=0.29\textwidth]{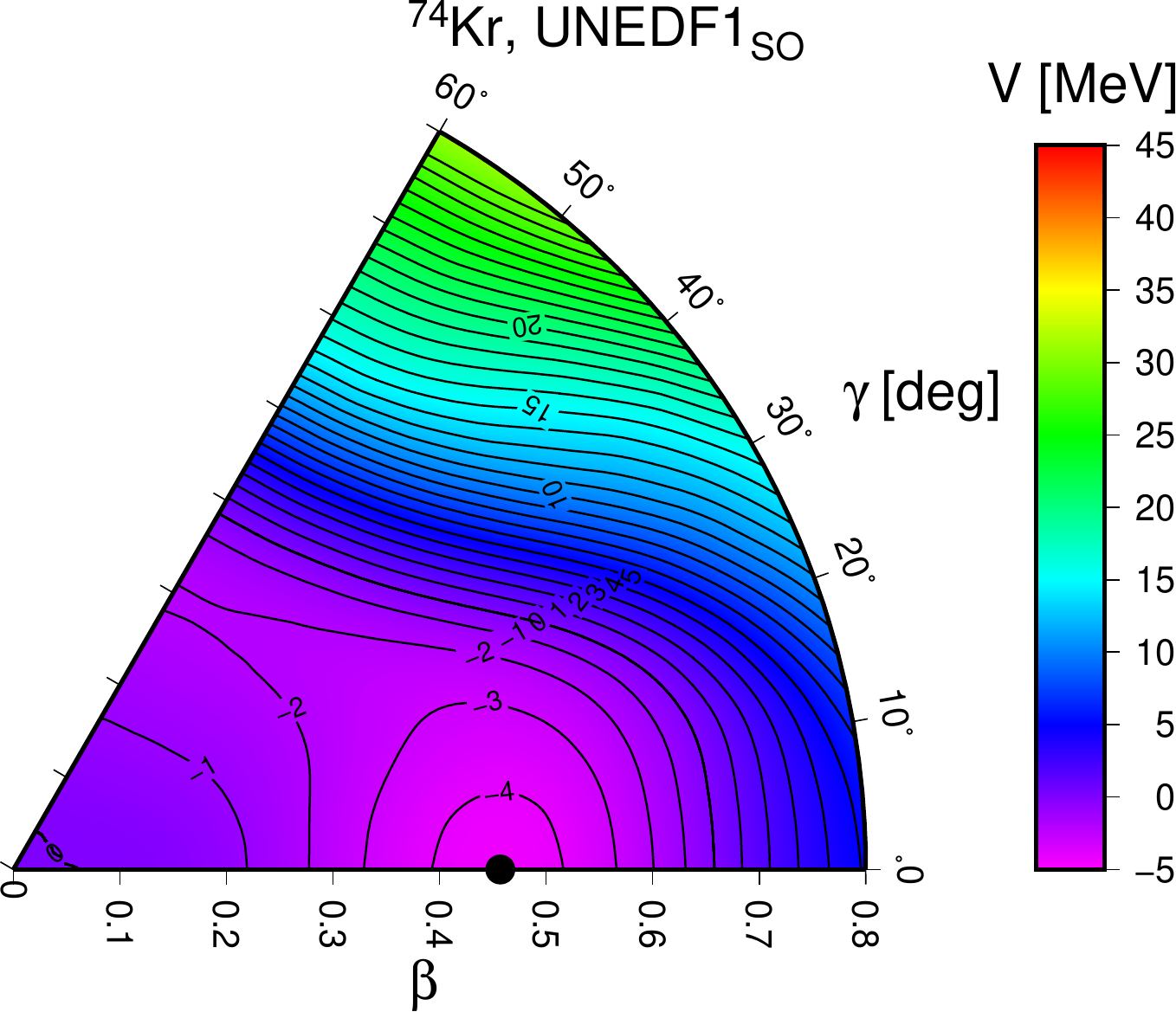}

\includegraphics[width=0.29\textwidth]{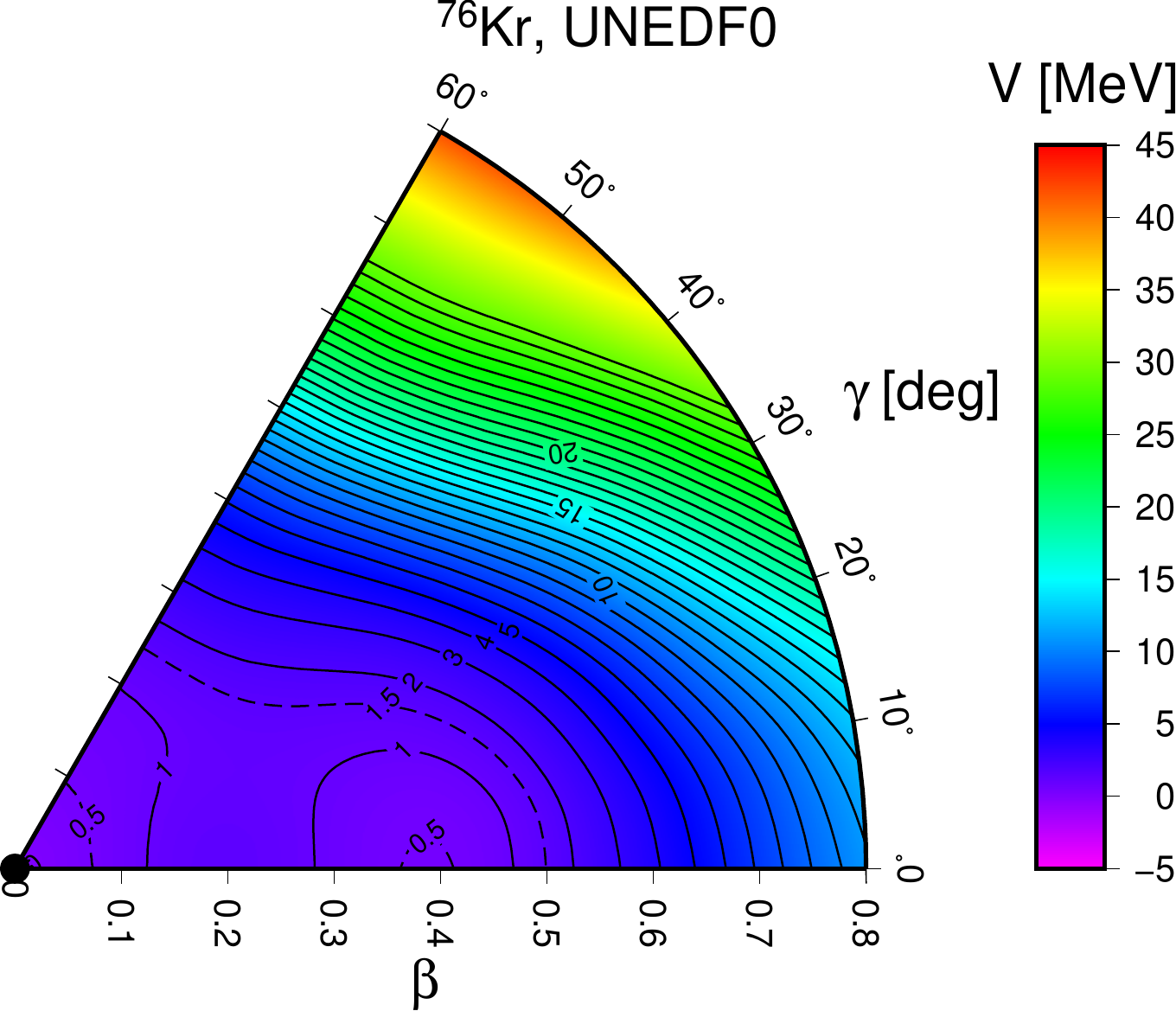}
\hfill
\includegraphics[width=0.29\textwidth]{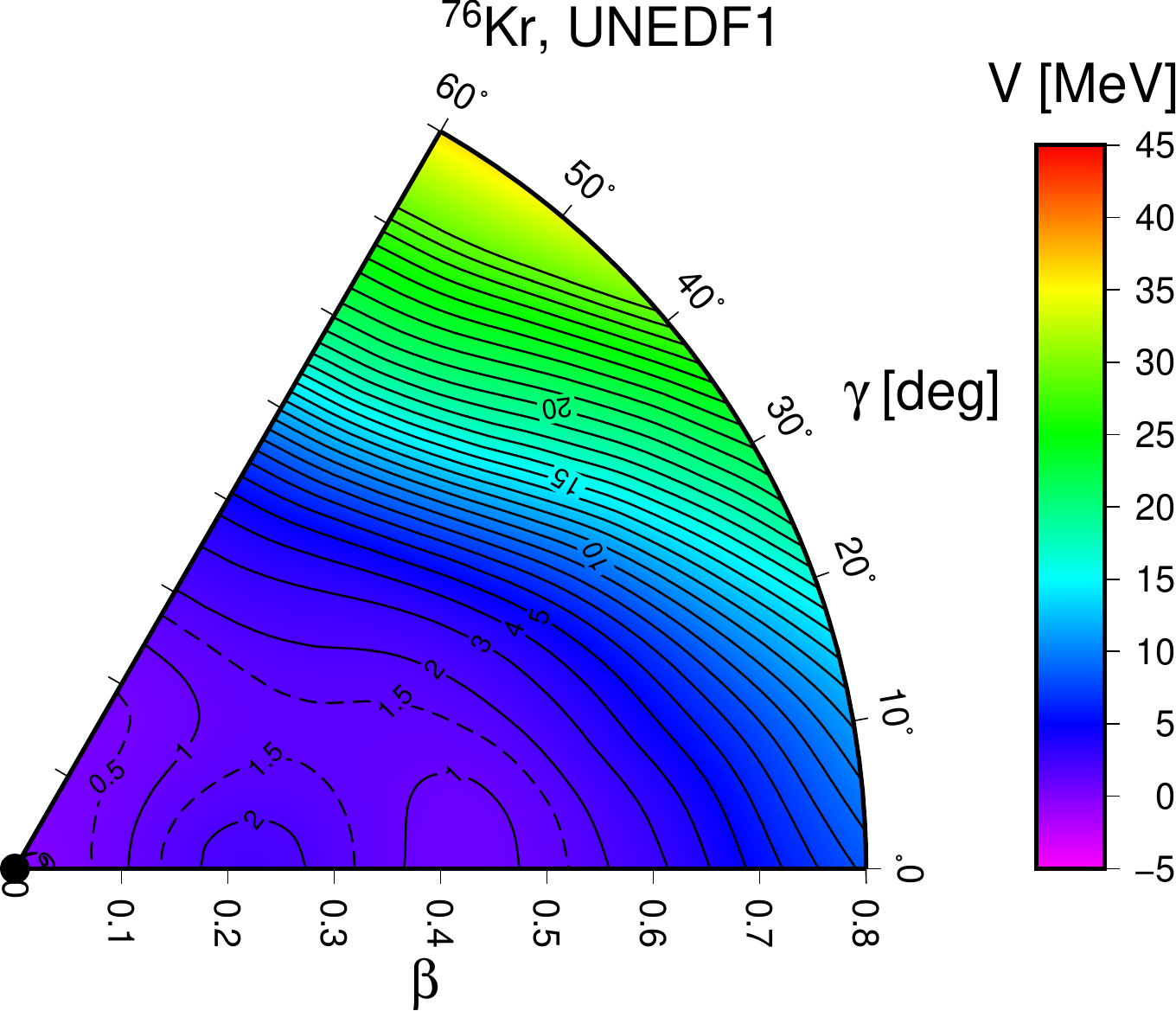}
\hfill
\includegraphics[width=0.29\textwidth]{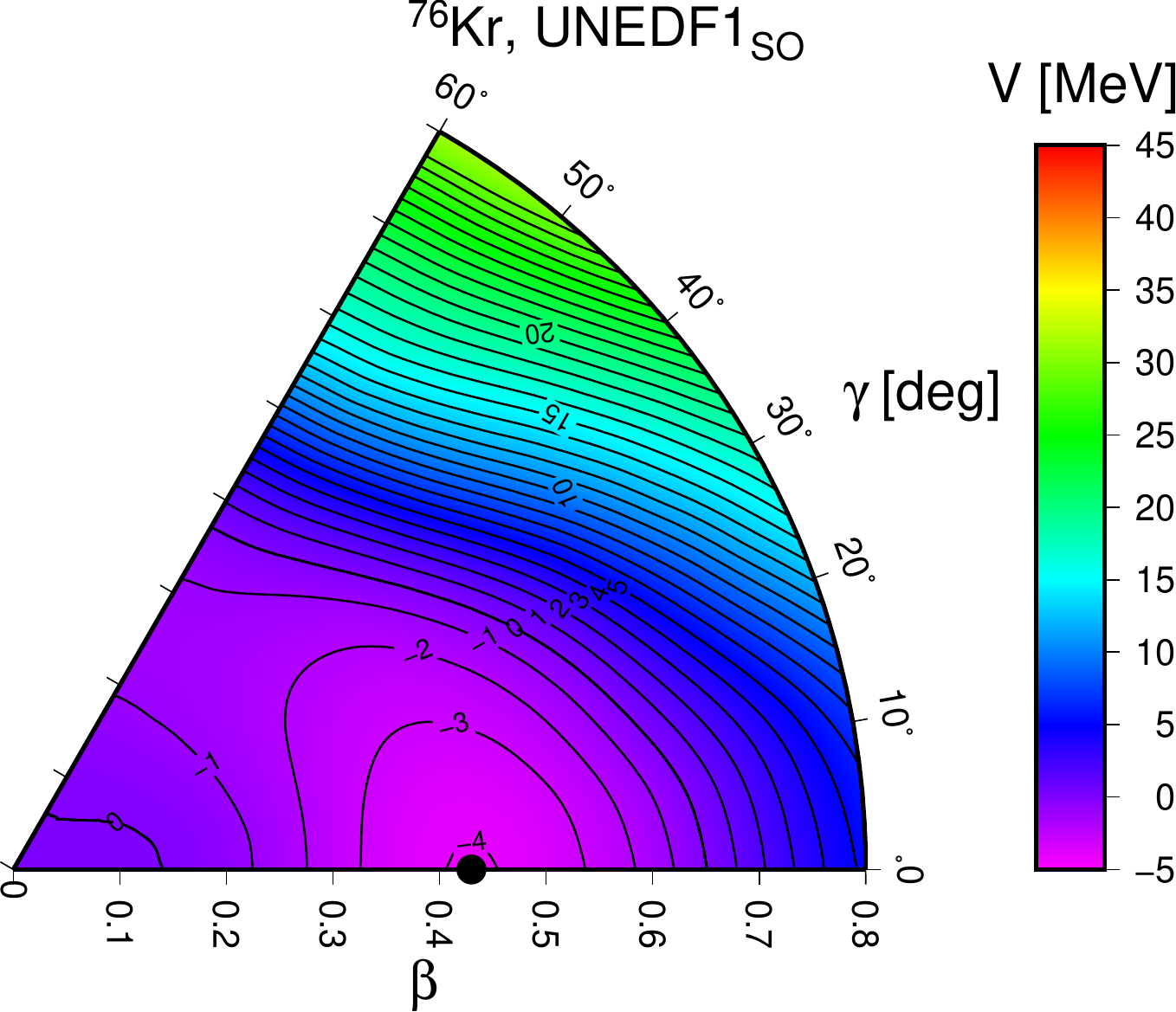}
\caption{(Colour Online) Potential energy surfaces for ${}^{72}$Kr, ${}^{74}$Kr and ${}^{76}$Kr obtained using UNEDF0, UNEDF1 and UNEDF1$_{\text{SO}}$ functionals.}
\label{Krypton PES}
\end{figure}

Having validated the overall quality of GBH methods based on Skyrme functionals, we now discuss in more detail the structure of neutron deficient Kr isotopes.
In Fig. \ref{Krypton PES}, we show the potential energy surfaces (PES) as obtained by solving constrained HFB equations at various points in the $\beta,\gamma$ plane.
The absolute energy minimum is indicated by a dot.
UNEDF0 predicts the $^{72}$Kr to be slightly oblate and the two other isotopes $^{74-76}$Kr spherical. It is important to notice here that the PES is quite flat along the $\gamma$ direction and we observe the appearance of secondary energy minima at prolate configurations at $\beta\approx0.4$.
For UNEDF1, we observe a more clear oblate minimum at $\beta\approx0.3$ in $^{72}$Kr, that reduces  to $\beta\approx0.1$ in $^{74}$Kr and becoming spherical in $^{76}$Kr. As seen for the UNEDF0 functional, also in this case we observe a well marked secondary minimum at $\beta\approx0.45$ for all three nuclei.
The case of UNEDF1$_{\text{SO}}$ is quite different, since the energy minimum on the PES is always on the prolate side for all 3 nuclei with $\beta\approx0.45$. In $^{72}$Kr we observe a secondary minimum in the oblate region.
In all cases, all functionals provide some softness toward triaxial configurations, although there are no real triaxial minima.

\begin{figure}[h!]
\centering
\begin{tikzpicture}
\def\PointSize{0.2pt}
\pgfplotsset{
   width = 15cm,
   height = 7.2cm, 
   compat = 1.9 ,
   yticklabel style = {/pgf/number format/.cd,fixed,fixed zerofill,precision=1} ,
   samples=3300,
}
\begin{axis}
[
xtick=\empty,
ytick distance = 0.5,
ytick pos = left,
xmin = 0,
xmax = 20,
ymin = -0.15,
ymax = 4.0,
ylabel={$E$ [MeV]}
]
\addplot[color=black] table{
5	-0.15
5	7
};

\addplot[color=black] table{
10	-0.15
10	7
};

\addplot[color=black] table{
15	-0.15
15	7
};

\addplot[color=gray!50] table{
1	0.0
19.5	0.0

1	0.70972
19.5	0.70972

1	1.32140
19.5	1.32140

1	2.1129
19.5	2.1129

1	3.1084
19.5	3.1084

3	0.6710
19.5	0.6710
};

\addplot[very thick] table{
1	0.0
2	0.0

1	0.70972
2	0.70972

1	1.32140
2	1.32140

1	2.1129
2	2.1129

1	3.1084
2	3.1084

3	0.6710
4	0.6710

6	0.0
7	0.0

6	0.7366201
7	0.7366201

6	1.4957642
7	1.4957642

6	2.3489726
7	2.3489726

6	3.3525258
7	3.3525258

8	1.0037833
9	1.0037833

11	0.0
12	0.0

11	0.6115925
12	0.6115925

11	1.3332155
12	1.3332155

11	2.1378639
12	2.1378639

11	3.0846505
12	3.0846505

13	0.8561730
14	0.8561730

16	0.0
17	0.0

16	0.2978999
17	0.2978999

16	0.8761063
17	0.8761063

16	1.7081949
17	1.7081949

16	2.7751702
17	2.7751702

18	1.3337678
19	1.3337678

};


\draw [color = black, ->] (axis cs:1.5,0.70972) -- (axis cs:1.5,0.0000000);

\node at (axis cs:1.5,0.35486) [fill = white] {\scriptsize $57\pm 8$};

\node (axis cs:1,0.0000000) [above] at (axis cs:2.3,-0.0500000) {\scriptsize $0^{+}_{1}$};
\node (axis cs:1,0.70972) [above] at (axis cs:2,0.70972) {\scriptsize $2^{+}_{1}$};
\node (axis cs:1,1.32140) [above] at (axis cs:2,1.32140) {\scriptsize $4^{+}_{1}$};
\node (axis cs:1,2.1129) [above] at (axis cs:2,2.1129) {\scriptsize $6^{+}_{1}$};
\node (axis cs:1,3.1084) [above] at (axis cs:2,3.1084) {\scriptsize $8^{+}_{1}$};
\node (axis cs:3,0.6710) [above] at (axis cs:4,0.6710) {\scriptsize $0^{+}_{2}$};

\draw [color = black, ->] (axis cs:6.5,0.7366201) -- (axis cs:6.5,0.0000000);
\draw [color = black, ->] (axis cs:6.5,1.4957642) -- (axis cs:6.5,0.7366201);
\draw [color = black, ->] (axis cs:6.5,2.3489726) -- (axis cs:6.5,1.4957642);
\draw [color = black, ->] (axis cs:6.5,3.3525258) -- (axis cs:6.5,2.3489726);

\node at (axis cs:6.5,0.36831005) [fill = white] {\scriptsize $31.385$};
\node at (axis cs:6.5,1.11619215) [fill = white] {\scriptsize $78.680$};
\node at (axis cs:6.5,1.9223684) [fill = white] {\scriptsize $130.816$};
\node at (axis cs:6.5,2.8507492) [fill = white] {\scriptsize $166.038$};

\node (axis cs:6,0.0000000) [red, above] at (axis cs:7,-0.0500000) {\scriptsize $0^{+}_{1}$};
\node (axis cs:6,0.7366201) [red, above] at (axis cs:7,0.6866201) {\scriptsize $2^{+}_{1}$};
\node (axis cs:6,1.4957642) [red, above] at (axis cs:7,1.4957642) {\scriptsize $4^{+}_{1}$};
\node (axis cs:6,2.3489726) [red, above] at (axis cs:7,2.3489726) {\scriptsize $6^{+}_{1}$};
\node (axis cs:6,3.3525258) [red, above] at (axis cs:7,3.3525258) {\scriptsize $8^{+}_{1}$};
\node (axis cs:8,1.0037833) [red, above] at (axis cs:9,1.0037833) {\scriptsize $0^{+}_{2}$};



\draw [color = black, ->] (axis cs:11.5,0.6115925) -- (axis cs:11.5,0.0000000);
\draw [color = black, ->] (axis cs:11.5,1.3332155) -- (axis cs:11.5,0.6115925);
\draw [color = black, ->] (axis cs:11.5,2.1378639) -- (axis cs:11.5,1.3332155);
\draw [color = black, ->] (axis cs:11.5,3.0846505) -- (axis cs:11.5,2.1378639);

\node at (axis cs:11.5,0.30579625) [fill = white] {\scriptsize $32.976$};
\node at (axis cs:11.5,0.972404) [fill = white] {\scriptsize $73.619$};
\node at (axis cs:11.5,1.7355397) [fill = white] {\scriptsize $135.044$};
\node at (axis cs:11.5,2.6112572) [fill = white] {\scriptsize $181.562$};

\node (axis cs:11,0.0000000) [violet,above] at (axis cs:12.5,-0.0500000) {\scriptsize $0^{+}_{1}$};
\node (axis cs:11,0.6115925) [violet,above] at (axis cs:12.5,0.6115925) {\scriptsize $2^{+}_{1}$};
\node (axis cs:11,1.3332155) [violet,above] at (axis cs:12,1.3332155) {\scriptsize $4^{+}_{1}$};
\node (axis cs:11,2.1378639) [violet,above] at (axis cs:12,2.1378639) {\scriptsize $6^{+}_{1}$};
\node (axis cs:11,3.0846505) [violet,above] at (axis cs:12,3.0846505) {\scriptsize $8^{+}_{1}$};
\node (axis cs:13,0.8561730) [violet,above] at (axis cs:14,0.8561730) {\scriptsize $0^{+}_{2}$};

\draw [color = black, ->] (axis cs:16.5,0.2978999) -- (axis cs:16.5,0.0000000);
\draw [color = black, ->] (axis cs:16.5,0.8761063) -- (axis cs:16.5,0.2978999);
\draw [color = black, ->] (axis cs:16.5,1.7081949) -- (axis cs:16.5,0.8761063);
\draw [color = black, ->] (axis cs:16.5,2.7751702) -- (axis cs:16.5,1.7081949);

\node at (axis cs:16.5,0.14894995) [left] {\scriptsize $97.764$};
\node at (axis cs:16.5,0.58700310) [fill = white] {\scriptsize $151.363$};
\node at (axis cs:16.5,1.29215060) [fill = white] {\scriptsize $179.369$};
\node at (axis cs:16.5,2.24168255) [fill = white] {\scriptsize $200.317$};

\node (axis cs:16,0.0000000) [blue,above] at (axis cs:17.7,-0.0500000) {\scriptsize $0^{+}_{1}$};
\node (axis cs:16,0.1978999) [blue,above] at (axis cs:17.7,0.2478999) {\scriptsize $2^{+}_{1}$};
\node (axis cs:16,0.8761063) [blue,above] at (axis cs:17,0.8761063) {\scriptsize $4^{+}_{1}$};
\node (axis cs:16,1.7081949) [blue,above] at (axis cs:17,1.7081949) {\scriptsize $6^{+}_{1}$};
\node (axis cs:16,2.7751702) [blue,above] at (axis cs:17,2.7751702) {\scriptsize $8^{+}_{1}$};
\node (axis cs:18,1.3337678) [blue,above] at (axis cs:19,1.3337678) {\scriptsize $0^{+}_{2}$};


\node [] at (axis cs:2.5,3.8) {\scriptsize Experimental Results};
\node [red] at (axis cs:7.5,3.8) {\scriptsize UNEDF0};
\node [violet] at (axis cs:12.5,3.8) {\scriptsize UNEDF1};
\node [blue] at (axis cs:17.5,3.8) {\scriptsize UNEDF1$_{\text{SO}}$};
\end{axis}
\end{tikzpicture}
\caption{(Colour Online) Shows the energy spectra for a number of low-lying states and their associated $B\left(E2\right)$ transition probabilities for ${}^{72}$Kr where we show the experimental results~\cite{DATA} (far left panel), UNEDF0 (left centre panel), UNEDF1 (right centre panel) and UNEDF1$_{\text{SO}}$ (far right panel).}
\label{Kr72 Spectra}
\end{figure}
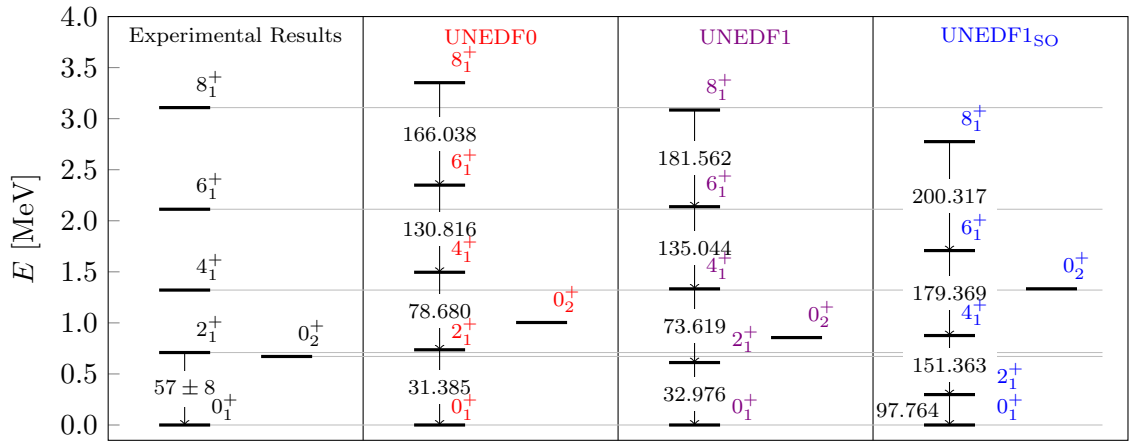

It is interesting to compare these PES with the one produced in Ref~\cite{gir09}. In this case the authors have used the Gogny D1S interaction~\cite{dec80}. A clear triaxial secondary minimum is observed in $^{72}$Kr, while here none of the used UNEDF functionals provide such a configuration.
The experimental spectrum of $^{72}$Kr Fig. \ref{Kr72 Spectra} is reproduced almost perfectly using the UNEDF1 functional, apart from a small deviation of $0.19$ MeV of the second $0^{+}_{2}$ compared to experimental findings.
The UNEDF0 and UNEDF1$_{\text{SO}}$ functionals give a slightly low (high) level density of the excited states.
Compared to the Gogny spectrum reported in Ref.~\cite{gir09}, we observe that the UNEDF functionals give on average a better description of the position of the energy states, although the $B\left(E2\right)$ are not well reproduced for the $B\left(E2:2^{+}_{1}\rightarrow 0^{+}_{1}\right)$ transition however are reproduced much better for the $B\left(E2:4^{+}_{1}\rightarrow 2^{+}_{1}\right)$ as illustrated by Fig \ref{2+ and 4+ BE2W of the Krypton Isotopic Chain}.
\section{Conclusion and Discussion}
We have analysed the evolution of the low-lying $2^{+}_{1}$ and $4^{+}_{1}$ states in the Kr isotopes using the Generalised Bohr Hamiltonian formalism.
By using 3 different Skyrme functionals, we have observed that the UNEDF0 and UNEDF1 functionals are able to provide, on average, quite a good description of both energy states and electromagnetic transitions.
We have studied in more detail 3 neutron deficient Kr isotopes, namely ${}^{72-76}$Kr. We have investigated the structure of their PES obtained with the various functionals and we have observed some softness against triaxial deformation in $^{72}$Kr, although we see no clear energy minimum as in the Gogny case.
By comparing the detailed structure of the full energy spectrum of $^{72}$Kr with the current experimental data, we have found that UNEDF1 gives a very nice reproduction of the spectrum.
In general we have observed that both UNEDF0 and UNEDF1 seem to provide a fairly good description of Kr isotopes. Similar conclusions were also found in Ref.~\cite{pro15} for the Xe isotopes. 
These results provide the motivation for a more systematic analysis of nuclear spectra using GBH together with these functionals to assess their quality in reproducing the data.

\ack
Supported by STFC Grant No.~ST/M006433/1, No.~ST/P003885/1 and by the Polish National Science Centre under contract No. 2018/31/B/ST2/02220

\section*{References}
\bibliographystyle{iopart-num}

\bibliography{biblio}

\end{document}